\documentclass[lettersize,journal]{IEEEtran}
\usepackage{amsmath,amsfonts}
\usepackage{algorithm}
\usepackage{array}
\usepackage{multirow}
\usepackage{amsmath}
\usepackage[caption=false,font=normalsize,labelfont=sf,textfont=sf]{subfig}
\usepackage{textcomp}
\usepackage{stfloats}
\usepackage{booktabs}
\usepackage{amssymb}
\usepackage{algpseudocode}
\usepackage{url}
\usepackage[hidelinks]{hyperref} 
\usepackage{verbatim}
\usepackage[table]{xcolor}
\usepackage{graphicx}
\usepackage{tabularx} 
\usepackage{color}
\usepackage{cite}

\begin{document}

\title{Generative Intent Prediction Agentic AI empowered Edge Service Function Chain Orchestration}

\author{Yan Sun, Shaoyong Guo*, Sai Huang,~\IEEEmembership{Senior Member,~IEEE,} Zhiyong Feng,~\IEEEmembership{Senior Member,~IEEE,} Feng Qi,\\ Xuesong Qiu,~\IEEEmembership{Senior Member,~IEEE}

\thanks{This work was supported by the National Natural Science Foundation of China (62321001). (*Corresponding author: Shaoyong Guo)}
\thanks{Yan Sun, Shaoyong Guo, Feng Qi and Xuesong Qiu are with the State Key Laboratory of Networking and Switching Technology, Beijing University of Posts and Telecommunications, Beijing, 100876, China (E-mail: \{sunyan79, syguo, qifeng, xsqiu\}@bupt.edu.cn).}
\thanks{Sai Huang and Zhiyong Feng are with the Key Laboratory of Universal Wireless
Communications, Beijing University of Posts and Telecommunications, Beijing 100876, China (E-mail: \{huangsai, fengzy\}@bupt.edu.cn).}

}

\markboth{Journal of \LaTeX\ Class Files,~Vol.~14, No.~8, August~2021}%
{Shell \MakeLowercase{\textit{et al.}}: A Sample Article Using IEEEtran.cls for IEEE Journals}

\maketitle

\begin{abstract}
With the development of artificial intelligence (AI), Agentic AI (AAI) based on large language models (LLMs) is gradually being applied to network management. However, in edge network environments, high user mobility and implicit service intents pose significant challenges to the passive and reactive management of traditional AAI. To address the limitations of existing approaches in handling dynamic demands and predicting users’ implicit intents, in this paper we propose an edge service function chain (SFC) orchestration framework empowered by a Generative Intent Prediction Agent (GIPA). Our GIPA aims to shift the paradigm from passive execution to proactive prediction and orchestration. First, we construct a multidimensional intent space that includes functional preferences, QoS sensitivity, and resource requirements, enabling the mapping from unstructured natural language to quantifiable physical resource demands. Second, to cope with the complexity and randomness of intent sequences, we design an intent prediction model based on a Generative Diffusion Model (GDM), which reconstructs users’ implicit intents from multidimensional context through a reverse denoising process. Finally, the predicted implicit intents are embedded as global prompts into the SFC orchestration model to guide the network in proactively and ahead-of-time optimizing SFC deployment strategies. Experiment results show that GIPA outperforms existing baseline methods in highly concurrent and highly dynamic scenarios.

\end{abstract}


\section{Introduction}
\IEEEPARstart{W}{ith} the development of artificial intelligence (AI), Agentic AI (AAI) built on large language models (LLMs) is being widely adopted in network management\cite{ref18,ref37}. By leveraging strong natural language understanding and contextual reasoning capabilities, AAI interpret users’ QoS intents expressed in natural language and make adaptive network management decisions, marking a shift toward a more advanced intent-driven networking paradigm\cite{ref19}. Especially in service function chain (SFC) orchestration, AAI can handle vague and unstructured instructions, autonomously decomposing high-level service goals into specific sequences of virtual network functions (VNFs) through logical reasoning\cite{ref20}.

However, in edge networks, user mobility and dynamic behavior introduce a large amount of implicit service intent, creating challenges for intent-driven network management\cite{ref21}. First, physical user movement often comes with immediate switches in service scenarios, which implicitly requires AAI to make predictive decisions such as service switching and task migration\cite{ref23}. If the agent only reacts passively, cold starts and migration delays can severely degrade user experience. Second, mobility increases user diversity, making it unrealistic to expect all users to express their intents clearly through a unified template. This requires AAI to infer users’ implicit intents and generate reasonable service orchestration strategies\cite{ref22}.

Therefore, to achieve truly intent-driven edge network management, it is urgent to design a new type of intent prediction AAI that helps the edge network perform proactive and predictive service orchestration. However, implementing an intent prediction agent in edge environments faces two main challenges:
\begin{itemize}
    \item \textbf{C1}: Complex mapping between implicit intents and service deployment decisions. Converting a vague and predictive intent into an optimal service deployment decision is highly challenging. This process involves not only translating semantic intent into physical resource requirements and node collaboration, but also handling the cooperation and competition between explicit and implicit intents\cite{ref24}. In addition, AAI must anticipate users’ future needs based on implicit intents and prepare in advance\cite{ref38}.
    \item \textbf{C2}: Complexity and randomness of intent sequences. Users’ intent evolution is not a simple linear sequence. Instead, it behaves as a stochastic process influenced by high-dimensional context such as time, location, and historical habits. Traditional prediction models, such as Recurrent Neural Network (RNN) and Long Short-Term Memory (LSTM), struggle to capture such complex distributions and often produce averaged predictions that miss critical implicit demands\cite{ref25}.
\end{itemize}

To address the above challenges, in this paper we propose a Generative Intent Prediction Agent (GIPA) empowered edge SFC orchestration framework. Unlike traditional passive and reactive management, GIPA aims to predict users’ implicit intents and respond in advance, enabling a paradigm shift from passive execution to proactive prediction and orchestration. 

Specifically, to tackle challenge C1, we build a mathematical model based on a multidimensional intent space. Instead of simple semantic matching, we construct an intent space that includes a functional preference vector, a QoS sensitivity vector, and a resource demand vector. Using the LLM embedded in GIPA, this modeling process quantifies vague and unstructured natural language instructions into measurable physical resource requirements and optimization objectives. To address challenge C2, we design an generative intent prediction model (GIPM) based on a Generative Diffusion Model (GDM)\cite{ref26}. This model leverages reverse denoising process to explore the latent distribution of intents, enabling accurate reconstruction of users’ implicit intents from noisy multidimensional context. This approach effectively overcomes the limitations of traditional prediction models when handling nonlinear intent evolution. In addition, we embed the predicted implicit intents into the SFC orchestration process. This mechanism allows orchestration strategies to move beyond current network state alone and proactively optimize VNF deployment based on predicted implicit demands, thereby maximizing service success rate and reducing end-to-end delay. The main contributions of this paper are as follows:
\begin{itemize}
    \item We propose an intent-driven proactive edge network management framework that integrates intent prediction based on heterogeneous AAI with an edge SFC orchestration mechanism. Through a lightweight domain classifier and a GIPA selection module, the framework efficiently interprets users’ natural-language intents under resource-constrained edge conditions, and perceives implicit service intents by analyzing context such as user demands and network state.
    \item We construct a multidimensional intent space and a prediction problem model. We decompose users’ implicit intents into vectors along three dimensions, including function, QoS, and resources, and build a context model that incorporates user mobility tendencies and network load conditions. This modeling process transforms a complex semantic understanding problem into a mathematical optimization problem with clear constraints, addressing the challenge of mapping abstract intents to concrete physical resource allocation.
    \item We design a generative prediction and intent-aware SFC orchestration algorithm. We use a GDM to handle the randomness of intent distributions and to generate implicit intent vectors. These vectors are then embedded as global prompts into the SFC orchestration model, guiding proactive service deployment and migration based on users’ implicit intents to maximize service continuity and QoS.
    \item We build an edge network simulation environment and compare GIPA against multiple baseline methods. Experimental results show that GIPA outperforms existing approaches in highly concurrent and highly dynamic scenarios. In particular, for service scenarios involving user mobility, GIPA achieves high service success rates and low-delay SFC orchestration.
\end{itemize}

The remainder of this paper is organized as follows. Section II reviews the related works. Section III introduces the design of GIPA framework and problem formulation. Section IV presents the design of generative predictive SFC orchestration model. Section V discusses the experimental results and analysis. Finally, Section VI concludes this paper.

\section{Related Works}
\subsection{Agentic AI}
With the maturation of LLMs and their reasoning frameworks, AAI is gradually replacing traditional automation scripts and emerging as a new paradigm for network management\cite{ref27}. Unlike rule-based systems, AAI can process vague instructions, autonomously plan task sequences, and invoke external tools, which makes it highly effective in tasks such as network fault diagnosis and automatic configuration generation\cite{ref28,ref39}. Existing research mainly focuses on enhancing reasoning chains and improving multi-agent collaboration, often relying on more sophisticated prompt engineering or hierarchical architectures to address large-scale network optimization problems. For example, the authors in \cite{ref40} are the first to propose an AAI framework that optimizes transmission strategies using a Mixture-of-Experts (MoE) approach. By employing a MoE-based proximal policy optimization method to address the problems formulated in AAI, the framework effectively improves both data transmission rates and the accuracy of AAI problem representation. The authors in \cite{ref1} proposed a multi-agent LLM system that decomposes a user's natural-language instruction into logical subtasks and converts them into configuration commands. The authors in \cite{ref2} introduced an autonomous management framework for mobile network services and resources based on AAI. Their framework employs a distributed multi-AAI system in which expert agents collaborate to interpret user requirements and deploy services using Infrastructure-as-Code (IaC) tools. In \cite{ref3}, the authors proposed a network–AI convergence architecture where multiple AAIs interact, learn collaboratively, and transfer knowledge to achieve autonomous solution discovery and dynamic environment adaptation.

However, mobile edge networks exhibit high dynamism and uncertainty. User service requests often vary rapidly with physical location and network conditions, and contain large amounts of implicit intent\cite{ref29}. Most existing AAI designs for network management still operate reactively, they respond only after network events occur or after users issue explicit instructions\cite{ref30}. To address these challenges, this paper proposes the GIPA framework. By continuously analyzing user behavior patterns and network dynamics, GIPA predicts users’ implicit demands and upcoming state changes, enabling a fundamental shift from passive execution to proactive management and better aligning with the high dynamism of edge environments.
\subsection{Intent-Driven Network}
Intent-driven networking (IDN) is an automated network management paradigm whose core idea is to allow users to describe high-level goals in semantic form, while the system automatically translates these intents into executable network policies\cite{ref31,ref42}. Unlike traditional networks that rely heavily on manual configuration, IDN operates through a closed-loop process that includes intent translation, policy generation, policy enforcement, and real-time monitoring\cite{ref32}. This enables the network to understand service requirements autonomously, adjust resource allocations dynamically, guarantee service quality, and adapt to environmental changes, thereby significantly reducing operational complexity. In essence, IDN shifts network behavior from configuration-driven to goal-driven operation.

With the adoption of LLMs in network management, the IDN paradigm has seen substantial progress. The authors in \cite{ref4} leveraged LLMs to automate intent-based management of shared networks. Their framework extracts declarative intents across business, service, and network planes to enable collaboration among network entities. The authors in \cite{ref5} proposed a generative intent abstraction framework based on LLMs. The framework uses LLMs and knowledge-enhanced prompts to dynamically generate adaptive intent graphs, improving the accuracy of complex intent translation. The authors in \cite{ref41} formulate an energy-efficiency maximization problem based on the transmitter’s intent, including power budget as well as the QoS requirements for information communication and energy harvesting. They further propose a PPO-based framework that jointly optimizes all variables, significantly enhancing energy efficiency and reducing decision-making time. In \cite{ref6}, the authors presented an LLM-based IDN solution for SDN, which seamlessly translates intents into programs by converting network intents into an intermediate representation and then invoking domain-specific code.

Despite these advances, existing IDN frameworks remain limited in handling latent user demands\cite{ref33}. In mobile networks, users’ true intents are often implicit and highly time-varying, causing network policies to satisfy only the literal meaning of user inputs rather than their actual service needs. To address this issue, the GIPA framework integrates a GDM-based implicit intent inference mechanism. Unlike simple semantic mapping, this mechanism processes multidimensional context with noise and uncertainty, and reconstructs users’ implicit intents accurately through a reverse denoising process.

\subsection{Service Function Chain}
SFC orchestration aims to dynamically map a sequence of VNFs onto underlying physical resources and is a key technology for ensuring edge service quality\cite{ref34,ref36}. Through centralized control and automated management, SFC orchestration enables flexible network service customization, efficient resource utilization, and rapid adaptation to changes in network conditions or service workloads, significantly enhancing network programmability and service quality\cite{ref35}. The authors in \cite{ref7} proposed a Two-Stage Reconfiguration method for SFC optimization, formalizing VNF migration and SFC reconstruction as a mathematical model to minimize either VNF migrations between nodes or the total number of core changes at each node. The authors in \cite{ref8} presented an SFC optimization method using adaptive multipath routing protection, which reduces reserved bandwidth on backup paths by distributing SFC traffic over disjoint operational paths across multiple failure zones. In \cite{ref9}, the authors proposed a dynamic multi-objective SFC placement method that partitions weight vectors into regions trained in parallel, optimizing SFC deployment by uncovering the relationship between weight vectors and solution locations.

In this paper, we propose an intent-driven predictive SFC orchestration model. By using the implicit intents predicted by GIPA as global conditions, our orchestration model is no longer limited to the current network state but considers users’ implicit or future demands. This enables the system to proactively deploy VNF instances on target nodes and plan optimal paths, ensuring service continuity.

\begin{figure*}[!t]
\centering
\includegraphics[width=\textwidth]{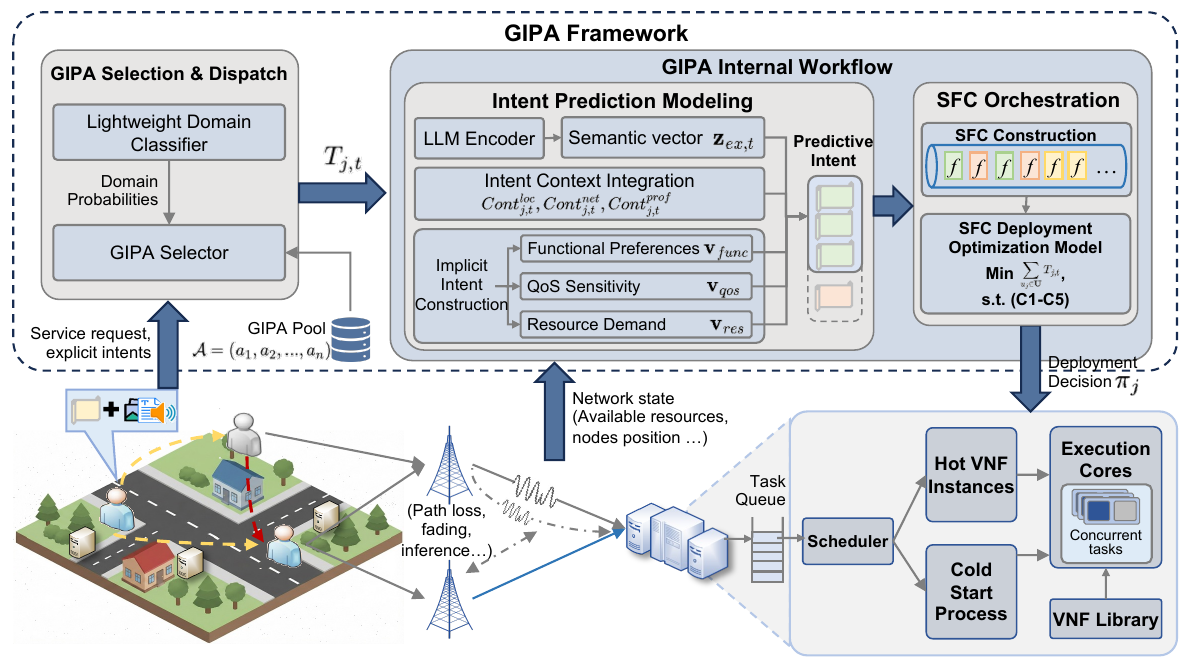}
\caption{Overview of GIPA empowered edge service function chain orchestration framework.}
\end{figure*}
\section{Problem Formulation}
In this section, we introduce the basic workflow of GIPA framework, as well as the problem formulation. The main notations are shown in Table I.
\begin{table}[tpb]
\renewcommand{\arraystretch}{1.5}
\caption{Main Notations\label{tab:table2}}
\centering
\begin{tabular}{p{2.5cm} p{5.5cm}}
\hline
$\textbf{Notation}$ & $\textbf{Description}$\\
\hline

$\mathbf{E},\mathbf{F},\mathbf{U}$ & The set of edge nodes, VNFs and users\\

$S_j$ & The ordered sequence of VNFs required to realize user $u_j$'s intents. \\

$T_{j,t}$ & The natural language intent of user $u_j$ at time $t$\\

$\mathbf{z}_{ex,t},\mathbf{z}_{im,t}$ & The explicit and implicit intent of user $u_j$ at time $t$ \\

$\mathbf{v}_{func,t},\mathbf{v}_{qos,t},\mathbf{v}_{res,t}$ & The function preference, QoS sensitivity and resource demand vector at time $t$ \\

$Cont_{j,t}$ & The intent space context\\

$a_n$ & The $n$-th AAI\\

$conf_{n,j}$ & The confidence score of agent an in translating intents within domain $d_j$\\

$\mathcal{N}_n^{OP},Tok_n^{OP}$ & The optimal number of parameters and the optimal number of training tokens given the computational budget\\

$e^{\mathcal{L}_n}$ & The intent translation accuracy of GIPA $a_n$\\

$Ad(a_n,T_{j,t})$ & The domain adaptability of GIPA $a_n$ to $T_{j,t}$\\

$C_f, M_f$  & The computing resources and data volume required by $f$\\

$\pi_{j,t}$  & The deployment nodes of $S_{j,t}$\\

$t_{j,i}^{queue},t_{j,i}^{proc}$ & The queuing and processing delay of $f_{j,i}$\\

$t_{j,i}^{comp},t_{j,i}^{comm}$ & The computation and communication delay of $f_{j,i}$\\

$T_{j,t}$ & The execution delay of SFC $S_{j,t}$\\

$\mathcal{Q}_t,\mathcal{H}_t$ & The queue of tasks in edge nodes and concurrent tasks at time $t$\\

$RP_{b,e}$ &  The reference signal received power between $b$ and $e$\\

$\chi_{e,b},\psi_{e,b}$ & The path loss and the shadow fading between $b$ and $e$\\

$\Phi$ & The signal to interference plus noise ratio \\

$Bw_{e,b}$ & The bandwidth resources allocated to $e$ from $b$ \\

$\mathcal{G}(\cdot)$ & The Gaussian cumulative distribution function \\

\hline
\end{tabular}
\end{table}

\subsection{Overview}
As shown in Fig. 1, in this paper we consider an edge network with m edge nodes, where the set of nodes is $\textbf{E} = \{e_1, e_2, ..., e_m\}$. Each edge node $e_i$ in \textbf{E} has limited computing resources, denoted as $r_i = \{C_i, M_i\}$, where $C_i$ and $M_i$ represent the CPU and memory resources of node $e_i$, respectively. The edge nodes host a library of VNFs $\textbf{F} = \{f_1, f_2, ..., f_k\}$. Assume the set of users in the network is $\textbf{U} = \{u_1, u_2, ..., u_n\}$. 

We represent user $u_j$’s intent as a vector $I_j$, which includes the user’s requirements for various QoS metrics. The SFC $S_j = \{f_{j,1}, f_{j,2}, ...\}$ represents the ordered sequence of VNFs required to realize these intents. The SFC deployment decision $\pi_j= \{e_{j,1}, e_{j,2}, ...\}$ indicates the nodes where each VNF in $S_j$ is deployed. GIPA consists of two main components: one is responsible for understanding users’ natural language intents and predicting implicit intents based on the LLM, and the other handles predictive SFC orchestration.
\subsection{Intent Prediction Modeling}
\subsubsection{Intent Space Construction}
To quantify user demands and capture their dynamic evolution, we first construct an intent space. Users’ explicit intent inputs are usually unstructured natural language text. We need to extract keywords and capture their latent semantic associations. Assume that at time $t$, user $u_j$’s explicit intent is $T_{j,t}$. We use the LLM in GIPA to map $T_{j,t}$ into a high-dimensional semantic vector $\mathbf{z}_{ex, t}$, which can be expressed as
\begin{equation}
    \mathbf{z}_{ex, t} = \mathcal{E}_{LLM}(T_{j,t}).
\end{equation}
The geometric properties of this vector ensure that semantically similar requests are close to each other in the space, providing a foundation for subsequent generalized implicit intent inference. Furthermore, we need to construct implicit intent vectors based on the semantic vectors. We decompose implicit intent into three subspaces:
\begin{itemize}
    \item Function preference vector $\mathbf{v}_{func}$: For the VNF library \textbf{F}, we define the functional preference vector as
    \begin{equation}
        \mathbf{v}_{func,t} = [p_{f_1}, p_{f_2}, \dots, p_{f_K}]^T \in [0,1]^K.
    \end{equation}
    Each element in the vector represents the probability that the user’s current service request depends on a specific network function. For example, for a video conferencing intent, the probabilities for video encoding and secure execution environment functions are relatively high.
    \item QoS sensitivity vector $\mathbf{v}_{qos}$: We model the user’s subjective importance of different QoS metrics as a QoS sensitivity vector:
    \begin{equation}
        \mathbf{v}_{qos, t} = [\omega_{Q_1}, \omega_{Q_2}, ..., \omega_{Q_q}]^T,
    \end{equation}
    where $\sum_{i=0}^q \omega_{Q_i} = 1$. This directly guides GIPA in selecting target nodes which can satisfy user's QoS requirements during SFC orchestration.
    \item Resource demand vector $\mathbf{v}_{res,t}$: This represents the implicit computing resource requirements and is used to assess the load imposed on each node:
    \begin{equation}
        \mathbf{v}_{res, t} = [c_t, m_t]^T.
    \end{equation}
\end{itemize}

Finally, we concatenate the above sub-vectors to obtain the implicit intent vector:
\begin{equation}
    \mathbf{z}_{im, t} = \text{Concat}(\mathbf{v}_{func, t}, \mathbf{v}_{qos, t}, \mathbf{v}_{res, t}) \in \mathbb{R}^{D_{im}}.
\end{equation}

In addition, users’ implicit service demands also depend on the current network state and the evolving state of the users themselves. We define this factor as the intent space context $Cont_{j,t}$. We consider $Cont_{j,t}$ to consist of three main components: user mobility, network state, and user profile, i.e., $Cont_{j,t} = [Cont_{j,t}^{loc}, Cont_{j,t}^{net}, Cont_{j,t}^{prof}]$.
\begin{itemize}
    \item User mobility context $Cont_{j,t}^{loc}$: User mobility directly determines which edge nodes can provide service coverage, thereby affecting SFC orchestration. We define a user’s trajectory within a time window $w$ as a sequence of edge nodes:
    \begin{equation}
        Cont_{j,t}^{loc} = \{ (e_{t-w}, \tau_{t-w}), \dots, (e_t, \tau_t) \},
    \end{equation}
    where $e_t \in E$ is the edge node the user is connected to, and $\tau_t$ is the residence time. By combining this context with user requests, we can predict when the user will move and proactively deploy VNFs at specific nodes.
    \item Network state context $Cont_{j,t}^{net}$: Obviously, the real-time load status of edge nodes affects SFC deployment decisions. We define the network state context as
    \begin{equation}
        Cont_{j,t}^{net} = \{ (r_{i,t}, B_t) \mid e_i \in \textbf{E} \},
    \end{equation}
    where $B_t$ represents the available bandwidth int the network.
    \item User profile context $Cont_{j,t}^{prof}$: We define the user profile context as a user preference vector, where each element indicates whether the user has a preference for a specific attribute, such as “prefers data-saving mode” or “prefers high-quality images.”
    \begin{equation}
        Cont_{j,t}^{prof} \in \mathcal{C}_{prof, i} \in \mathbb{R}^{d_{prof}},
    \end{equation}
    where $d_prof$ represents the number of preference attributes.
\end{itemize}

Based on the above discussion, we can represent the implicit intent prediction process as
\begin{equation}
    \mathbf{z}_{im, t} = \mathcal{M}(\mathbf{z}_{ex, t}, Cont_{j,t}),
\end{equation}
where $\mathcal{M}$ represents the implicit intent mapping method. Ideally, this would be a direct mapping function, but due to the ambiguity of natural language and the stochasticity of user behavior, $\mathcal{M}$ is highly nonlinear. Therefore, we reformulate the problem as a conditional probability distribution
\begin{equation}
    p(\mathbf{z}_{im, t} \mid \mathbf{z}_{ex, t}, Cont_{j,t}).
\end{equation}
The learning process of this conditional probability distribution is detailed in Section IV. Based on the implicit intent, we can construct the SFC $S_{j,t}$.
\subsubsection{GIPA Selection}
Considering the performance limitations of edge nodes, it is difficult to deploy large-scale LLMs within GIPA at the edge, making it challenging to balance generality and performance\cite{ref10}. In this paper, we consider deploying a set of heterogeneous GIPA agents in the edge network, $\mathcal{A} = \{a_1, a_2, \dots, a_N\}$, where each GIPA specializes in intent translation for different domains. When a user issues a service request, we first match the user with the GIPA agent expected to achieve the best performance for intent translation.

First, we divide network service intents into $D$ typical functional domains, denoted as $\mathcal{D} = {d_1, \dots, d_D}$. For any GIPA $a_n \in \mathcal{A}$, we define its capability feature vector as $\mathbf{conf}n = [conf{n,1}, conf_{n,2}, \dots, conf_{n,D}]^T \in [0,1]^D$, where $conf_{n,j}$ represents the confidence score of agent $a_n$ in translating intents within domain $d_j$. This parameter can be obtained through offline evaluation or online feedback.

For a natural language intent input $T_{j, t}$ submitted by user $u_j$, we first estimate its probability distribution over domains using a lightweight classifier, producing the vector $\mathbf{w}{j,t} = [w{j,t,1}, \dots, w_{j,t,D}]^T$, where $\sum w_{j,t,d} = 1$. The domain adaptability of GIPA $a_n$ to $T_{j, t}$ can then be expressed as
\begin{equation}
    Ad(a_n,T_{j, t})= \mathbf{conf}_n^T \cdot \mathbf{w}_t = \sum_{d=1}^D conf_{n,d} \cdot w_{j,t,d}.
\end{equation}

The intent translation performance depends not only on the GIPA agent’s expertise in the domain but also on the scale of the LLM model within the GIPA and its pretraining quality. According to the LLM scaling laws \cite{ref11}, the pretraining performance of an LLM depends on the computational budget $C^{train}$ during training. Therefore, the intent translation pretraining performance of the LLM within GIPA can be expressed as \cite{ref11}
\begin{equation}
    \mathcal{L}_n=\mathcal{L}_0+\frac{\xi}{(N_n^{OP})^\alpha_1}+\frac{\kappa}{(Tok_n^{OP})^\alpha_2},
\end{equation}
\begin{equation}
    \mathcal{N}_{n}^{OP} = \left(\frac{\alpha_{1}\xi}{\alpha_{2}\kappa}\right)^{\frac{1}{\alpha_{1}+\alpha_{2}}} \left(\frac{C^{train}}{6}\right)^{\frac{\alpha_{2}}{\alpha_{1}+\alpha_{2}}},
\end{equation}
\begin{equation}
    Tok_{n}^{OP} = \left(\frac{\alpha_{2}\kappa}{\alpha_{1}\xi}\right)^{\frac{1}{\alpha_{1}+\alpha_{2}}} \left(\frac{C^train}{6}\right)^{\frac{\alpha_{1}}{\alpha_{1}+\alpha_{2}}}, 
\end{equation}
where $\mathcal{L}_0$ represents the loss of an ideal generative process, which can be considered as 0. $\mathcal{N}_{n}^{OP}$ and $Tok_{n}^{OP}$ are the optimal number of parameters and the optimal number of training tokens given the computational budget $C^{train}$. The term $\frac{\xi}{(\mathcal{N}_n^{OP})^{\alpha_1}}$ represents the performance loss due to limited model parameters, and $\frac{\kappa}{(Tok_n^{OP})^{\alpha_2}}$ represents the performance loss due to limited training data. According to the LLM scaling laws \cite{ref11}, $\alpha_1$, $\alpha_2$, $\xi$, and $\kappa$ are set to 0.34, 0.28, 406.4, and 410.7, respectively. Based on this, the intent translation accuracy of GIPA $a_n$ can be expressed as $e^{-\mathcal{L}_n}$. Furthermore, we can derive the accuracy of $a_n$ in translating the natural language intent $T_{j, t}$ as
\begin{equation}
    Acc(a_n,T_{j, t})=Ad(a_n,T_{j, t})\cdot e^{-\mathcal{L}_n}.
\end{equation}

Based on the above analysis, when a user issues a service request, we select the GIPA agent with the highest translation accuracy for that specific request to interpret the intent.

\subsection{QoS Modeling}
In edge scenarios, the execution delay of an SFC is the most critical metric for measuring QoS, primarily consisting of computation delay and communication delay.
\subsubsection{Computation Delay}
Computation delay mainly includes queuing delay and task processing delay. Assume that at time $t$, user $u_j$ submits a service request and the implicit intent $\mathbf{z}{im, t}$ is obtained. Based on this, we construct the SFC $S_{j,t} = \{f_{j,1}, f_{j,2}, ...\}$, where $f_{j,i} \in \textbf{F}$. We denote $r_{j,i} = \{C_{j,i}, M_{j,i}\}$ as the computing resources and data volume required by $f_{ji}$. $\pi_{j,t} = \{e_{j,1}, e_{j,2}, ...\}$ represents the deployment nodes of each subtask in $S_{j,t}$. 

Let $\zeta_{f_{j,i}}$ denote the CPU frequency (cycles per second) of node $e_{j,i}$ and $\eta_{f_{j,i}}$ the number of CPU cores allocated to $f_{j,i}$. After $f_{j,i}$ is assigned to node $e_{j,i}$, it enters the waiting queue. Assume the current queue is $\mathcal{Q}_t$ and the set of concurrent tasks at this moment is $\mathcal{H}_t$. We first calculate $t_{j,i}'$, the time $f_{j,i}$ waits for concurrent tasks to complete. Considering that concurrent tasks are simultaneously assigned to $e_{j,i}$ and their execution order is random, we use the average execution time of concurrent tasks to represent the waiting time for a single task.
\begin{equation}
    t_{j,i}’=\frac{1}{|\mathcal{H}_t|}\sum_{f\in \mathcal{H}_t}\frac{C_f M_f}{\zeta_f \eta_f}.
\end{equation}
The execution of $f_{j,i}$ depends on the instantiation of the corresponding VNF on node $e_{ji}$. When a corresponding instantiated VNF exists on $e_{j,i}$, $e_{j,i}$ can directly warm-start the VNF container to execute $f_{j,i}$. Conversely, when the corresponding instantiated VNF does not exist on $e_{j,i}$, it requires re-instantiating a new VNF container to execute $f_{j,i}$. We refer to this process as a cold start. Since a cold start involves operations such as VNF instance creation, resource configuration, and image distribution, the delay of a cold start is significantly higher than that of a warm start \cite{ref12}. In this paper, we focus solely on the impact of cold start delay on task execution delay. We consider a scenario where the node is fully loaded, meaning that multiple VNFs are instantiated in $e_{j,i}$ and all computing resources are fully occupied. Assuming the cold start delay is $t_{cold}$, if the VNF instance corresponding to $f_{j,i}$ currently exists in $e_{j,i}$, and the set of tasks of the same type in the queue is denoted as $\mathcal{Q}'$, then the queuing delay of $f_{ji}$ is defined as the waiting time for the completion of other tasks of the same type:
\begin{equation}
    t_{j,i}^{queue}=\sum_{f \in \mathcal{Q}’}\frac{C_f M_f}{\zeta_f \eta_f}.
\end{equation}

If the VNF instance corresponding to $f_{j,i}$ does not currently exist in $e_{j,i}$, then it is necessary to wait for the completion of all tasks preceding $f_{j,i}$ in the queue. Assuming that there are $n$ types of tasks without corresponding VNF instances, the queuing delay of $f_{j,i}$ is:
\begin{equation}
    t_{j,i}^{queue}=\sum_{f \in \mathcal{Q}}\frac{C_f M_f}{\zeta_f \eta_f}+(n+1)t_{cold}.
\end{equation}
The processing delay of $f_{j,i}$ is
\begin{equation}
    t_{j,i}^{proc}=\frac{C_{j,i} M_{j,i}}{\zeta_{f_{j,i}}  \eta_{f_{j,i}}}.
\end{equation}
Then the total computation delay of $f_{j,i}$ is
\begin{equation}
    t_{j,i}^{comp}=t_{j,i}’+t_{j,i}^{queue}+t_{j,i}^{proc}.
\end{equation}

\subsubsection{Communication Delay}
In edge networks, a task's communication delay is not only determined by the allocated bandwidth but is also affected by factors such as channel interference and encoding schemes. Assume that the edge node $e$ executing $f_{j,i}$ communicates with the user via base station $b$, and the bandwidth allocated by $b$ to $e$ is $Bw_{e,b}$. Furthermore, assume there are $\lambda$ ray clusters between $b$ and $e$, with $\gamma$ rays in each cluster, and the signal transmission power of $b$ is $P$. Then, the Reference Signal Received Power between $e_{j,i}$ and $b$ is given by \cite{ref13}:
\begin{equation}
    RP_{b,e}=\chi_{e,b} \psi_{e,b} (|\alpha_0|^2+\sum_{a=1}^\lambda \sum_{b=1}^\gamma|\alpha_{a,b}|^2)P,
\end{equation}
where $\chi_{e,b}$ represents the path loss between $e$ and $b$, and $\psi_{e,b}$ represents the shadow fading. $\alpha_0$ and $\alpha_{a,b}$ denote the power of the line-of-sight (LoS) path and the non-line-of-sight (NLoS) path, respectively. Inter-cell interference primarily depends on the six cells with the highest Reference Signal Received Power \cite{ref13}. Assuming the set of base stations in the current network is $\mathbf{B}$, then the communication interference power between $b$ and $e$ is:
\begin{equation}
    P_{e,b}^{inf}=\sum_{\beta \in \mathbf{B}/{b}}max^{(6)}(RP_{e,\beta}).
\end{equation}
Then, the Signal-to-Interference-plus-Noise Ratio (SINR) between $e$ and $b$ is:
\begin{equation}
    \Phi_{e,b}=\frac{RP_{b,e}}{\sigma^2+P_{e,b}^{inf}},
\end{equation}
where $\sigma$ represents the noise power. Assuming the coding scheme used for transmitting data related to $f_{j,i}$ has an error rate of $\epsilon$, then according to the transmission rate formula under finite block length \cite{ref14}, the data transmission rate between $e$ and $b$ is
\begin{equation}
    \begin{array}{l}
    \nu=C-\sqrt{\frac{cd}{bl}Q^{-1}(\epsilon)}\\
=Bw_{e,b}log_2(\Phi_{e,b})-\mathcal{G}^{-1}(\epsilon)\sqrt{\frac{\Phi_{e,b}}{2bl}\cdot \frac{\Phi_{e,b}+2}{(\Phi_{e,b}+1)^2}log_2^2e},
\end{array}
\end{equation}
where $cd$ represents the channel dispersion, $bl$ represents the block length, and $\mathcal{G}(\cdot)$ represents the Gaussian cumulative distribution function (CDF). Assuming the amount of data transmitted from $e$ to $b$ is $M_{e,b}$, then the communication delay is:
\begin{equation}
    t_{e,b}=M_{e,b}/\nu.
\end{equation}

In this paper, we only consider the communication delay between the edge node and the base station, because the communication delay between the user and the base station depends on the bandwidth allocated by the operator, a factor we deem to be outside the management scope of the method presented in this paper. Therefore, the communication delay of $f_{j,i}$ is:
\begin{equation}
    t_{j,i}^{com}=t_{e,b}+t_{b,e}.
\end{equation}
Ultimately, the total execution delay of $f_{j,i}$ is derived as
\begin{equation}
    t_{j,i}=t_{j,i}^{comp}+t_{j,i}^{com}.
\end{equation}
Furthermore, the total delay of $S_{jt}$ is derived as
\begin{equation}
    T_{j,t}=\max_{f \in S_{j,t}}t_f.
\end{equation}

Our objective is to minimize the total execution delay by adjusting the SFC deployment and resource allocation. We define a binary decision variable $x_{j,k,i} \in \{0, 1\}$. If the $k$-th Virtual Network Function (VNF) $f_{j,k}$ in the SFC $S_{j,t}$ of user $u_j$ is deployed on edge node $e_i$, $x_{j,k,i}=1$; otherwise, $x_{j,k,i}=0$. Therefore, we formulate the following optimization problem:
\begin{equation}
\begin{aligned}
\text{\textbf{OP}}: & \min \sum_{u_j \in \mathbf{U}} T_{j,t} \\
\text{s.t.:} \quad & \text{(C1):} \sum_{e_i \in \mathbf{E}} x_{j,k,i} = 1, \quad \forall u_j \in \mathbf{U}, \forall f_{j,k} \in S_{j,t} \\
& \text{(C2):} \sum_{u_j \in \mathbf{U}} \sum_{f_{j,k} \in S_{j,t}} x_{j,k,i} \cdot C_{f_{j,k}} \le C_i, \quad \forall e_i \in \mathbf{E} \\
& \text{(C3):} \sum_{u_j \in \mathbf{U}} \sum_{f_{j,k} \in S_{j,t}} x_{j,k,i} \cdot M_{f_{j,k}} \le M_i, \quad \forall e_i \in \mathbf{E} \\
& \text{(C4):} \sum_{u_j \in \mathbf{U}} \sum_{f_{j,k} \in S_{j,t}} x_{j,k,i} \cdot b_{f_{j,k}} \le Bw_{e_i, b}, \quad \forall e_i \in \mathbf{E} \\
& \text{(C5):} \sum_{e \in \mathbf{E}} Bw_{e,b} \le Bw_b, \quad \forall b \in \mathbf{B}
\end{aligned}
\end{equation}
Where Constraint C1 ensures that every sub-function in each SFC can only be deployed on a single node. Constraints C2 and C3 ensure that the allocated computing and storage resources do not exceed the node's total capacity. C4 mandates that the bandwidth occupied by each sub-function cannot exceed the total bandwidth allocated by the base station to the node. C5 ensures that the bandwidth allocated to the nodes cannot exceed the base station's total bandwidth capacity.
\section{Generative Predictive SFC Orchestration Model}
In this section, we introduce the architecture and training process of the GPSO Model. As shown in Fig. 2, the GPSO Model is composed of the GIPM, which is responsible for Implicit Intent Prediction, and the PSOM, which is responsible for predictive SFC orchestration.
\begin{figure*}[!t]
\centering
\includegraphics[width=\textwidth]{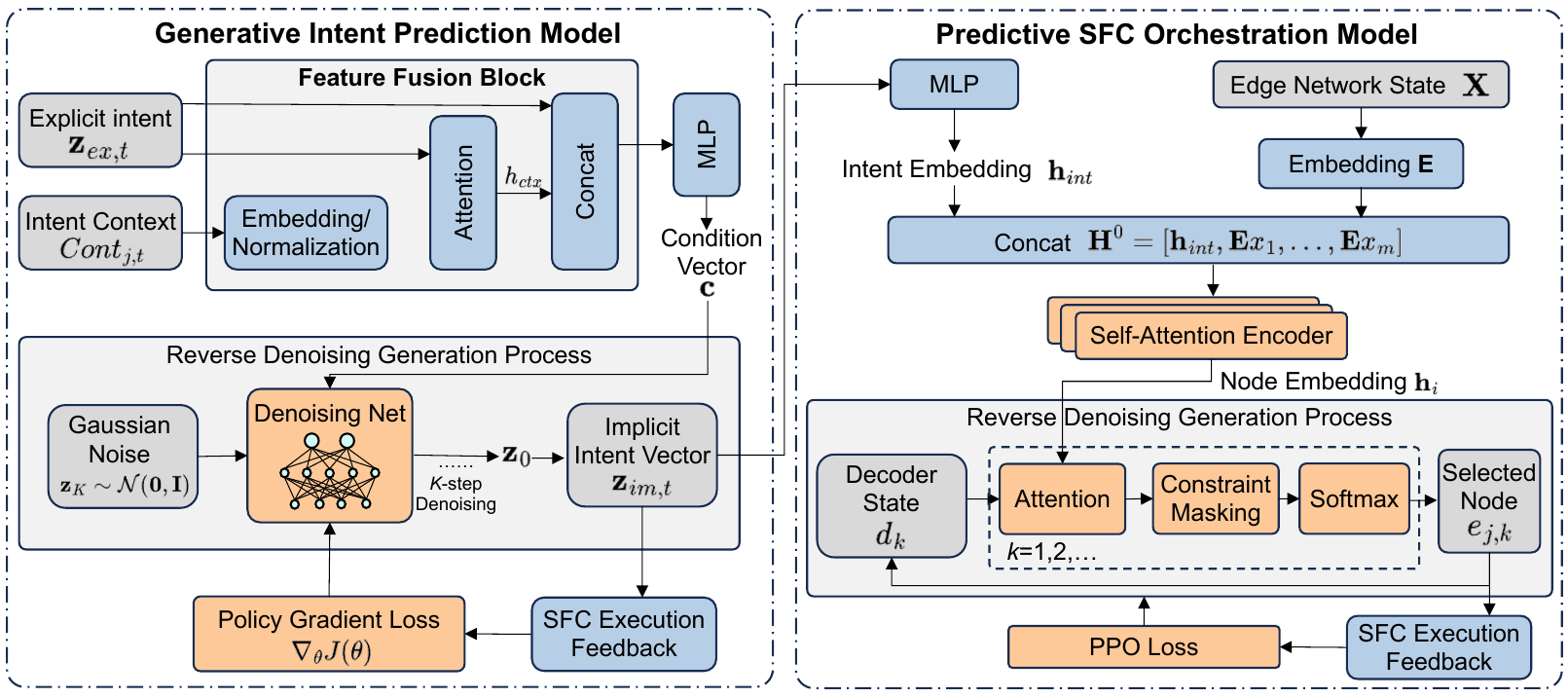}
\caption{Overview of Generative Predictive SFC Orchestration Model.}
\end{figure*}
\subsection{Generative Intent Prediction Model}
The core task of the GIPM is to construct a mapping from explicit intent and multi-dimensional context to implicit intent. As described in Section III, this is a conditional probability distribution modeling problem: $p(\mathbf{z}_{im, t} \mid \mathbf{z}_{ex, t}, Cont_{j,t})$. Given the complex nonlinear coupling relationships among heterogeneous features in the implicit intent vector $\mathbf{z}_{im, t}$ and the strong randomness of user intent, we utilize the GDM as the core of intent prediction in this paper. Leveraging its powerful exploration capability, it offers a greater advantage in capturing implicit variable relationships \cite{ref15}.

We train the GIPM using the Deep Reinforcement Learning (DRL) paradigm, and its decision-making process can be modeled as a Markov Decision Process (MDP). The state space is $s_t^G=[\mathbf{z}_{ex, t},Cont_{j,t}]$, which includes the explicit intent of the user's current service request and the context of the service request. The action space is $a_t^G= \mathbf{z}_{im, t} = [\mathbf{v}_{func}, \mathbf{v}_{qos}, \mathbf{v}_{res}]$, which comprises the function dependency probability, QoS weights, and resource requirements. 

Due to the lack of ground-truth labels, we evaluate the accuracy of implicit intent generation through the SFC execution feedback. The model is penalized if the generated resource requirements are too low, leading to task failure, or if inappropriate QoS weight allocation results in unfulfilled key performance indicators. The reward function is defined as 
\begin{equation}
    r_t^G=\frac{T_{max}-T_{j,t}}{T_{max}-T_{tar}}-\lambda Prob_{fail},
\end{equation}
where $T_{tar}$ represents the user's target delay, $T_{max}$ represents the maximum possible delay for a single service, $Prob_{fail}$ represents the service failure probability, and $\lambda$ represents the weight assigned to the failure probability in the reward.

In order to guide the generation process of the diffusion model, we need to map the heterogeneous input information into a unified latent space to serve as the conditional vector. We utilize a Multi-Layer Perceptron (MLP) as the feature fusion network. First, the discrete context features (such as node IDs in $Cont_{j,t}^{loc}$) are embedded, and continuous features (such as resource quantities in $Cont_{j,t}^{net}$) are normalized. Subsequently, an Attention mechanism is employed to dynamically capture the correlation strength between the semantic vector and the context information:
\begin{equation}
    \mathbf{h}_{ctx} = \text{Attention}(\mathbf{z}_{ex, t} \mathbf{W}_Q, Cont_{j,t} \mathbf{W}_K, Cont_{j,t} \mathbf{W}_V).
\end{equation}
Finally, the conditional vector $\mathbf{c}$ is defined as the fused feature representation:
\begin{equation}
    \mathbf{c} = \text{MLP}(\text{Concat}(\mathbf{z}_{ex, t}, \mathbf{h}_{ctx})).
\end{equation}

The vector $\mathbf{c}$ will be injected into the denoising network to guide the reconstruction of the implicit intent. It is worth noting that the traditional DDPM (Denoising Diffusion Probabilistic Models) includes a forward diffusion and a reverse generation process. However, considering that user intent prediction is not a labeled task and there is a lack of a dataset showing the optimal intent prediction results, our model does not include the forward diffusion process. Our model only includes the reverse denoising generation process, which is given by
\begin{equation}
    \begin{array}{l}
{\pi  }({a_t^G}|{s_t^G}) = {p }({\mathbf{z}_{0:N}}|\mathbf{c})\\
 = {{\cal N}}({\mathbf{z}_N};0,{\bf{I}})\prod\limits_{n = 1}^N {{p }({\mathbf{z}_{n - 1}}|{\mathbf{z}_n},\mathbf{c})},
 \end{array}
\end{equation}
where $\pi (a_t^G|s_t^G)$ represents the action selection policy of the GIPM, $\mathbf{I}$ is the identity matrix, and $N$ is the number of denoising steps.

The reverse denoising process starts from standard Gaussian noise $\mathbf{z}_K \sim \mathcal{N}(\mathbf{0}, \mathbf{I})$ and yields the action $\mathbf{z}_{im, t}$ after $K$ denoising steps. Each step is given by:
\begin{equation}
    \mathbf{z}_{k-1} = \frac{1}{\sqrt{\alpha_k}} \left( \mathbf{z}_k - \frac{1-\alpha_k}{\sqrt{1-\bar{\alpha}_k}} \epsilon_\theta(\mathbf{z}_k, k, \mathbf{c}) \right) + \sigma_k \mathbf{z},
\end{equation}
where $\alpha_k = 1 - \beta_k$ represents the signal retention coefficient, which measures how much original information is retained during the current diffusion step. $\beta_k$ represents the predefined noise variance, which is typically a very small number. $\sigma_k$ represents the standard deviation of the random noise added during the reverse generation process, used to measure the randomness in the generation process.

To train this diffusion policy in the absence of labels, our training objective is no longer to minimize the reconstruction error (MSE) but to maximize the expected reward $J(\theta) = \mathbb{E}_{\pi_\theta}[r_t^{G}]$. We utilize policy gradient methods to update the denoising network parameters $\theta$, thereby increasing the probability density of high-reward intent samples in the generated distribution. The training algorithm of GIPM is shown in Algorithm 1.
\begin{equation}
    \nabla_\theta J(\theta) \approx \mathbb{E} \left[ \sum_{k=1}^K \nabla_\theta \log p_\theta(\mathbf{z}_{k-1}|\mathbf{z}_k, \mathbf{c}) \cdot Q(\mathbf{z}_{im}, s_t) \right].
\end{equation}

\begin{algorithm}[h]
\caption{GIPM Training}
\begin{algorithmic}
\State \textbf{Input:} Edge Network State $\mathcal{X}$, User Explicit Intent $T_{j,t}$, Intent Context $Cont_{j,t}$.
\State \textbf{Initialize:} $\text{GIPM}$ Denoising Network $\theta_{G}$.
\State \textbf{Hyperparameters:} $N_{episodes}$, $T_{max}$, $K$ (denoising steps), $\lambda$ (failure penalty), learning rate.
\State \textbf{Reward Function:} $r_{t} = \frac{T_{max} - T_{j,t}}{T_{max} - T_{tar}} - \lambda Prob_{fail}$

\For{episode = 1 to $N_{episodes}$}
    \State Reset environment and network state $\mathcal{X}$.
    \For{timestep $t$ = 1 to $T_{max}$}
        \State $z_{ex,t} \gets \mathcal{E}_{LLM}(T_{j,t})$
        \State $h_{ctx} \gets \text{Attention}(z_{ex,t}, Cont_{j,t})$
        \State $c \gets \text{MLP}(\text{Concat}(z_{ex,t}, h_{ctx}))$
        \State $z_{K} \sim \mathcal{N}(0, I)$
        \State $z_{im,t} \gets \text{Reverse Denoising}(z_{K}, c, K)$
        
        \State $S_{j,t} \gets \text{SFC\_Construction}(z_{im,t})$
        \State $\pi_{j} \gets \text{SFC\_Deployment}(z_{im,t}, \mathcal{X}, \theta_{P})$
        \State $T_{j,t}, Prob_{fail} \gets \text{Execute\_SFC}(\pi_{j})$
        \State $r_{t} \gets \text{Reward}(T_{j,t}, Prob_{fail})$
        
        \State \textbf{Update $\theta_{G}$:} $\theta_{G} \gets \theta_{G} + \alpha \nabla_{\theta}J(\theta)$
    \EndFor
\EndFor
\State \textbf{Return:} Trained $\text{GIPM}$ $\theta_{G}$.
\end{algorithmic}
\end{algorithm}

\subsection{Predictive SFC Orchestration Model}
Based on the implicit intent, we obtain the required SFC $S_{j,t}$ for user $j$. We need to make an appropriate deployment decision for $S_{j,t}$ based on the current network environment and the user's implicit intent. First, we construct the state of the $m$ edge nodes as a feature matrix $\mathbf{X} \in \mathbb{R}^{m \times {3}}$. For each node $e_i$, its feature vector $x_i$ includes the remaining computing, storage, and bandwidth resources:
\begin{equation}
    x_i = [C_{rem, i}, M_{rem, i}, Bw_{rem, i}].
\end{equation}

We treat the implicit intent $\mathbf{z}_{im, t}$ output by the GIPM as a global prompt. Using a MLP, we map it to an intent embedding vector $\mathbf{h}_{int}$ that shares the same dimension as the node features.
\begin{equation}
    \mathbf{h}_{int} = \text{MLP}(\mathbf{z}_{im, t}).
\end{equation}

In order for the encoder to simultaneously perceive the network topology and the user intent, we concatenate the intent embedding before the node feature sequence, forming the enhanced input sequence $\mathbf{H}^0 = [\mathbf{h}_{int}, \mathbf{E}x_1, \dots, \mathbf{E}x_m]$. Subsequently, the sequence is encoded through $L$ layers of Multi-Head Self-Attention. The computation at the $l$-th layer is given by:
\begin{equation}
    \mathbf{Q} = \mathbf{H}^{l-1}\mathbf{W}_Q, \quad \mathbf{K} = \mathbf{H}^{l-1}\mathbf{W}_K, \quad \mathbf{V} = \mathbf{H}^{l-1}\mathbf{W}_V,
\end{equation}
\begin{equation}
    \mathbf{H}^l = \text{Softmax}\left(\frac{\mathbf{Q}\mathbf{K}^T}{\sqrt{d_k}}\right)\mathbf{V} + \mathbf{H}^{l-1}.
\end{equation}

Finally, we obtain the output embedding $h_i$ for node $i$. The deployment of the SFC is a sequence decision process, and we use an attention decoder to obtain the node selection probability for each service. Assume that in step $k$, we need to select a deployment node for the $k$-th VNF $f_{j,k}$ in the SFC. Let $d_k$ be the decoder's hidden state at time $k$ (updated from the embedding of the previously selected node and the current VNF requirement $r_{j,k}$). We calculate the correlation score $u_{i,k}$ between $d_k$ and all edge node embeddings $\mathbf{h}_i$:
\begin{equation}
    u_{i,k} = v^T \tanh(\mathbf{W}_{ref} \mathbf{h}_i + \mathbf{W}_q d_k),
\end{equation}
where $v_t$, $\mathbf{W}_{ref}$, and $\mathbf{W}_q$ represent learnable neural network parameters. Then, in order for the deployment to satisfy Constraints C1-C5, we define the masking function:
\begin{equation}
    \mathcal{M}_{i,k} = \begin{cases} 0, & \text{if C1-C5 are satisfied}  \\ -\infty, & \text{otherwise} \end{cases}.
\end{equation}

Finally, the probability of deploying the $k$-th VNF to node $e_i$ is given by
\begin{equation}
    p(e_{j,k} = e_i \mid s_t) = \text{Softmax}(u_{i,k} + \mathcal{M}_{i,k}).
\end{equation}

We employ the Proximal Policy Optimization (PPO) algorithm to train the PSOM. We use the same reward function in the PSOM as in the GIPM to ensure consistency between the deployment policy and the intent prediction performance, $r_t^P=\frac{T_{max}-T_{j,t}}{T_{max}-T_{tar}}-\lambda Prob_{fail}$. The training algorithm of GIPM is shown in Algorithm 2.

\begin{algorithm}[h]
\caption{PSOM Training}
\begin{algorithmic}
\State \textbf{Input:} Edge Network State $\mathcal{X}$, User Explicit Intent $T_{j,t}$, Intent Context $Cont_{j,t}$.
\State \textbf{Initialize:} $\text{PSOM}$ Attention Decoder $\theta_{P}$, Trained $\text{GIPM}$ $\theta_{G}$.
\State \textbf{Hyperparameters:} $N_{episodes}$, $T_{max}$, PPO parameters.
\State \textbf{Reward Function:} $r_{t} = \frac{T_{max} - T_{j,t}}{T_{max} - T_{tar}} - \lambda Prob_{fail}$

\For{episode = 1 to $N_{episodes}$}
    \State Reset environment and network state $\mathcal{X}$.
    \For{timestep $t$ = 1 to $T_{max}$}
        \State $z_{im,t} \gets \text{GIPM\_Predict}(T_{j,t}, Cont_{j,t}, \theta_{G})$
        \State $S_{j,t} \gets \text{SFC\_Construction}(z_{im,t})$
        
        \State $h_{int} \gets \text{MLP}(z_{im,t})$
        \State $H^{L} \gets \text{Encoder}(\text{Concat}(h_{int}, \mathcal{X}))$
        \State $\pi_{j} \gets \{\,\}$
        \State $d_{0} \gets \text{InitialDecoderState}$

        \For{$k=1$ to $|S_{j,t}|$}
            \State $f_{j,k} \gets S_{j,t}[k]$
            \State $d_{k} \gets \text{UpdateDecoderState}(d_{k-1}, r_{j,k})$
            \State $u_{i,k} \gets \text{AttentionScore}(H^{L}, d_{k})$
            \State $\mathcal{M}_{i,k} \gets \text{ConstraintMask}(e_{i}, \text{C1-C5})$
            \State $p(e_{j,k}=e_{i}) \gets \text{Softmax}(u_{i,k} + \mathcal{M}_{i,k})$
            \State $e_{j,k} \sim p(e_{j,k})$
            \State $\pi_{j} \gets \pi_{j} \cup \{e_{j,k}\}$
        \EndFor
        
        \State $T_{j,t}, Prob_{fail} \gets \text{Execute\_SFC}(\pi_{j})$
        \State $r_{t} \gets \text{Reward}(T_{j,t}, Prob_{fail})$
        
        \State \textbf{Update $\theta_{P}$:} $\theta_{P} \gets \text{PPO\_Update}(\theta_{P}, r_{t})$
    \EndFor
\EndFor
\State \textbf{Return:} Trained $\text{PSOM}$ $\theta_{P}$.
\end{algorithmic}
\end{algorithm}

\section{Numerical Results}
\textbf{Training Setup}: We implement and deploy the GIPA on a server equipped with an AMD EPYC 7763 CPU and 4 GeForce RTX 4090 (24GB) GPUs. We use the DeepSeek-R1 (7B) model\footnote{\url{https://ollama.com/library/deepseek-r1}.} as the Large Language Model (LLM) in the GIPA for understanding user intent. The GPSO model was trained using PyTorch 2.5.0. Due to the lack of a natural language intent dataset, we utilized Gemini 3 Pro\footnote{\url{https://gemini.google.com}} to generate 4,000 natural language instructions covering various scenarios (e.g., ``HD video calling," ``low-delay autonomous driving," ``real-time data processing"). During the GPSO training process, we trained the GIPM model first, followed by the PSOM model. Considering that the GIPM training relies on feedback from the downstream SFC deployment performance, we used brute-force search to find the optimal SFC deployment during the GIPM training phase. After the GIPM training achieved convergence under this condition, we began training the PSOM using the GIPM output.

\textbf{Environment Setup}: We built an edge network simulation environment using Python. $M=20$ Edge Nodes and $B=5$ Base Stations (BSs) are randomly distributed within a $5 \text{km} \times 5 \text{km}$ area. The CPU capacity $C_i$ for each edge node $i$ follows a uniform distribution of $[20, 50]$ GHz, the storage $M_i$ follows a uniform distribution of $[100, 300]$ GB, and the bandwidth $Bw_{e,b}$ follows a uniform distribution of $[100, 500]$ Mbps. We set the VNF library to contain 10 different types of functions. The length of the SFC requested by each user ranges from $3$ to $6$ VNFs.

\textbf{Baseline Setup}: To validate the performance of GIPA, we compare it with the following three baseline service orchestration methods:
\begin{itemize}
    \item IMAAC\cite{ref16}: An edge service scheduling model trained under a Multi-Agent Actor-Critic framework.
    \item MTRL\cite{ref17}: An SFC orchestration model that exploits the shared characteristics among tasks to improve generalization capability and learning efficiency.
    \item Best Effort (BE): A simple method that allocates each VNF to the nearest node with abundant resources.
\end{itemize}
\subsection{Performance Comparison}
First, we compare the overall performance of GIPA with that of the baseline methods. We primarily observe two metrics: service execution delay and success rate. During the tests, we gradually increase the number of concurrent tasks to observe the performance comparison among the different methods.
\begin{figure}[h]
\centering
\subfloat[]{\includegraphics[width=2.5in]{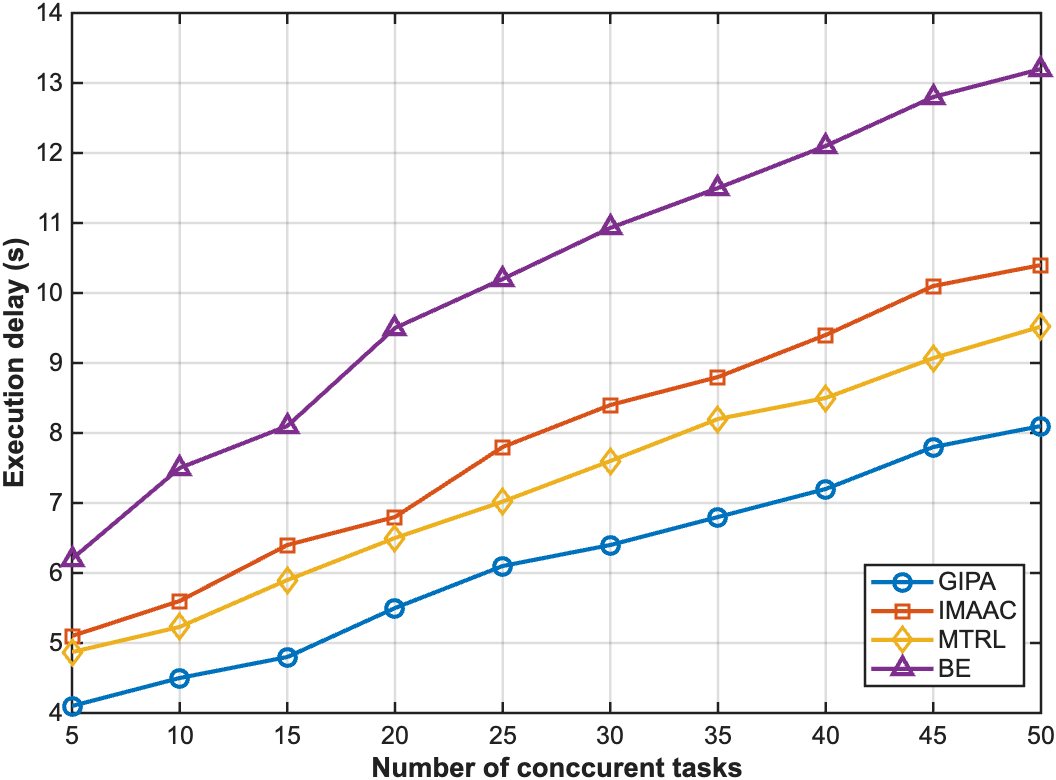}%
\label{9a}}\\
\subfloat[]{\includegraphics[width=2.5in]{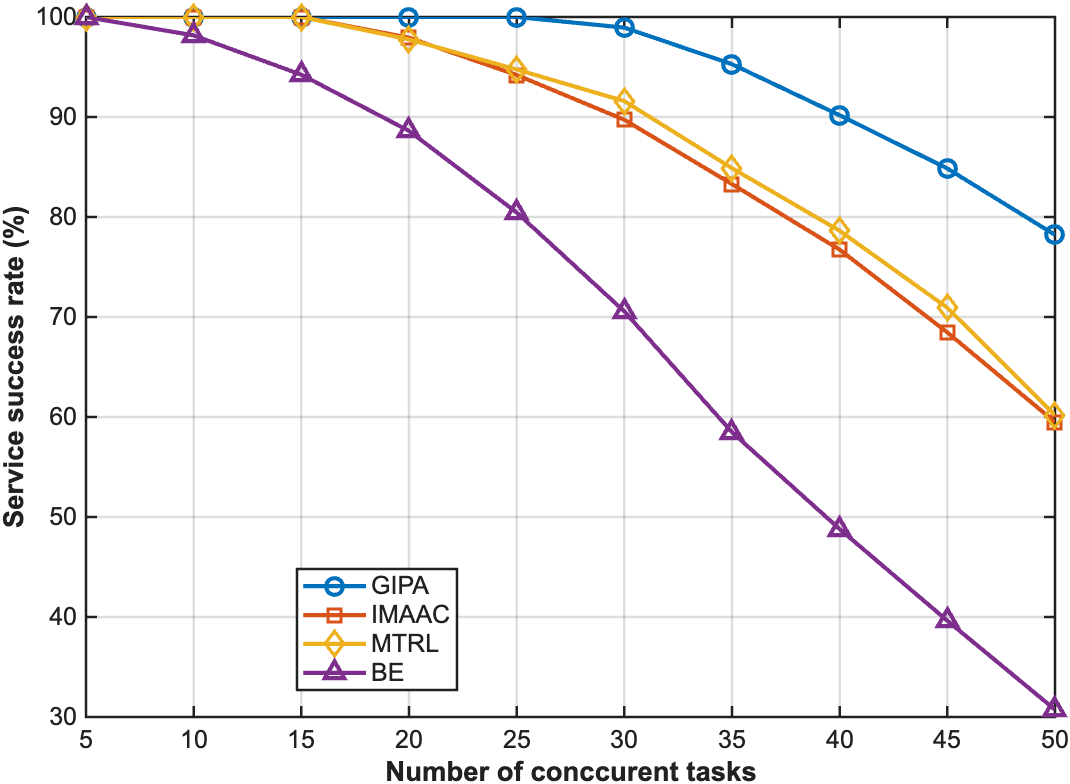}%
\label{9b}}\\
\caption{Performance comparison between different methods. (a) Execution delay comparison. (b) Success rate comparison.}
\label{fig9}
\end{figure}

As shown in Fig. 3(a), as the number of concurrent tasks increases from 5 to 50, the average execution delay of all methods shows an upward trend, which is due to increased competition for edge node computing resources, resulting in longer queuing delays. However, GIPA consistently maintains the lowest execution delay. BE performs the worst because it selects nodes based only on distance, neglecting the current load status and bandwidth limitations of the nodes. This leads to tasks being stacked on hotspot nodes, resulting in long queuing delays. Although IMAAC and MTRL optimize scheduling through reinforcement learning, they primarily make reactive decisions based on the current network state. In contrast, GIPA exhibits the gentlest increase in delay. This is primarily attributed to its internal GIPM, which can proactively predict the user's implicit intent, enabling the early deployment of VNFs that can either accelerate the current service execution or be deployed ahead of time in areas where the mobile user is expected to appear, thereby reducing delay.

As shown in Fig. 3(b), as the number of concurrent tasks increases, the task success rate of the baseline methods gradually declines, while GIPA consistently maintains a higher service success rate. Service failure often stems from disruptions caused by user mobility in the edge network. Several baseline methods are reactive, only triggering migration when the link is broken or the signal attenuates. This not only increases delay but is also highly prone to task failure. Leveraging the user request context, GIPA is able to interpret the trend of an impending handover area from the implicit intent. The PSOM utilizes this prediction to proactively deploy the SFC on nodes ahead of the user's mobility path. This proactive predictive orchestration strategy circumvents service interruptions caused by mobility, thus achieving the highest service success rate in dynamic edge environments.

To intuitively demonstrate the effect of GIPA's implicit intent prediction on improving QoS, we specifically tested the performance of several methods when orchestrating navigation services. Navigation services often inherently contain the implicit intent of user movement, and timely prediction of user movement followed by service migration or the pre-deployment of related services near the target node can effectively improve QoS.
\begin{figure}[h]
\centering
\includegraphics[width=2.5in]{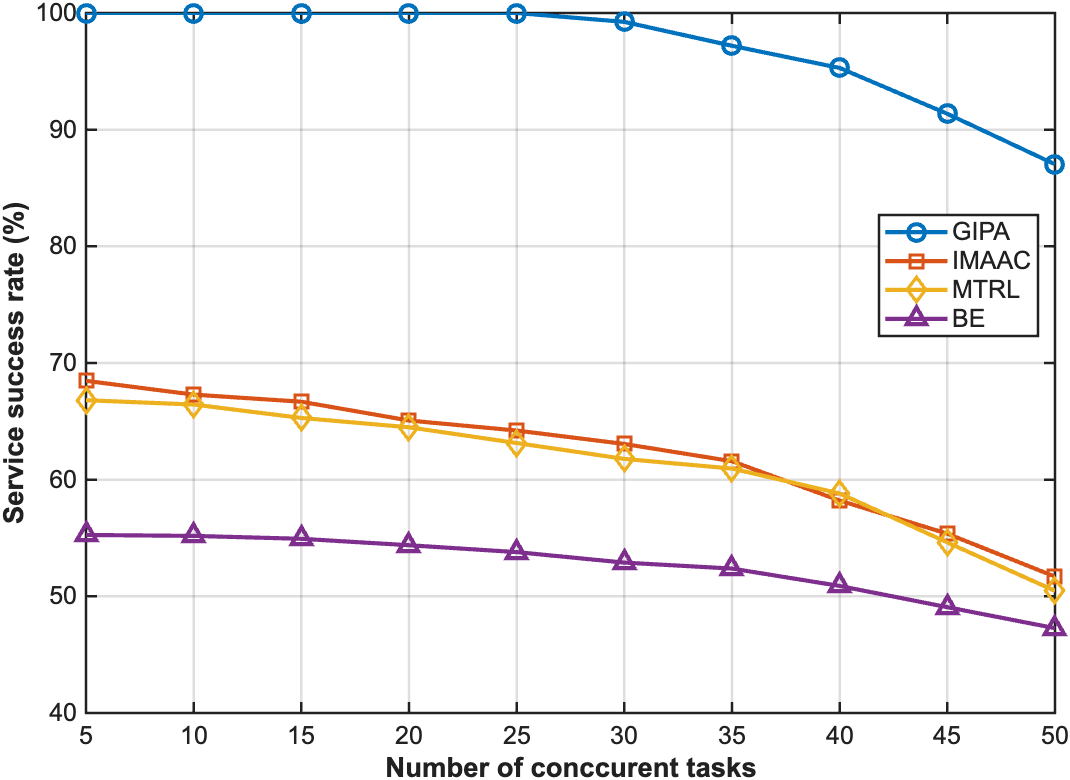}%
\caption{Performance of several methods when orchestrating navigation services.}
\end{figure}

As shown in Fig. 4, compared to the success rate of mixed services in Fig. 3(b), the success rate of the dedicated navigation service demonstrates a larger gap between the baseline methods and GIPA. Because BE, IMAAC, and MTRL lack the ability to model user trajectory context, they can only react passively to location changes. This results in services often requiring high-delay cross-node migration and cold start processes after user movement, consequently lowering the service success rate. It should be noted that under high concurrency, the success rates of all methods show an improvement compared to the results in Figure 3(b). This is because the navigation services occupy fewer computing resources, leading to an enhanced capability of the system to handle high concurrency.
\subsection{Ablation Experiment}
In this section, we use ablation experiments to demonstrate the efficacy of the modules designed in this paper. 
\subsubsection{The effect of implicit intent prediction}
\begin{figure}[h]
\centering
\includegraphics[width=2.5in]{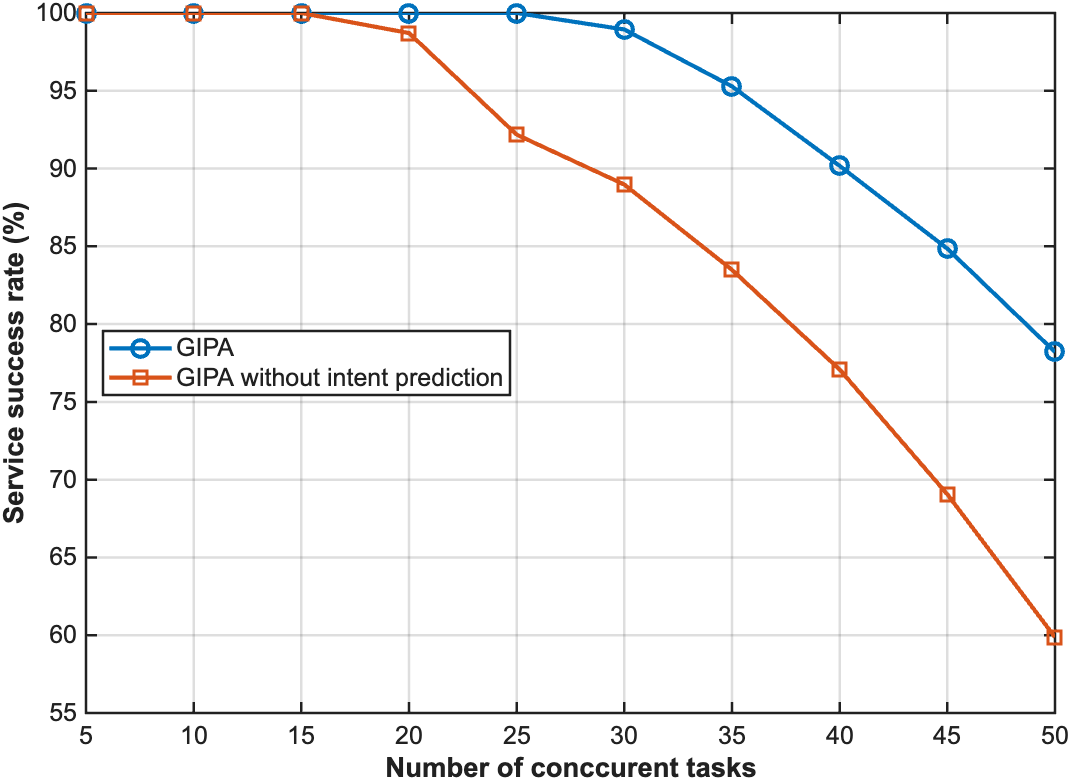}%
\caption{Performance of several methods when orchestrating navigation services.}
\end{figure}
To validate the role of implicit intent, we trained a GIPA variant without the GIPM and observed its performance compared to the standard GIPA. As shown in Fig. 5, the service success rate of GIPA lacking intent prediction significantly decreases under high concurrency. Although the LLM itself can understand user intent, the lack of the intent prediction module makes it difficult to predict the user's implicit intent, resulting in lower QoS. For example, for an autonomous driving request, if the prediction of the implicit, high-priority QoS of low delay is missing, the orchestration model might treat it as a regular computing task, leading to key metrics not being satisfied and consequently resulting in service failure.
\subsubsection{The effect of GDM in intent prediction}
\begin{figure}[h]
\centering
\includegraphics[width=2.5in]{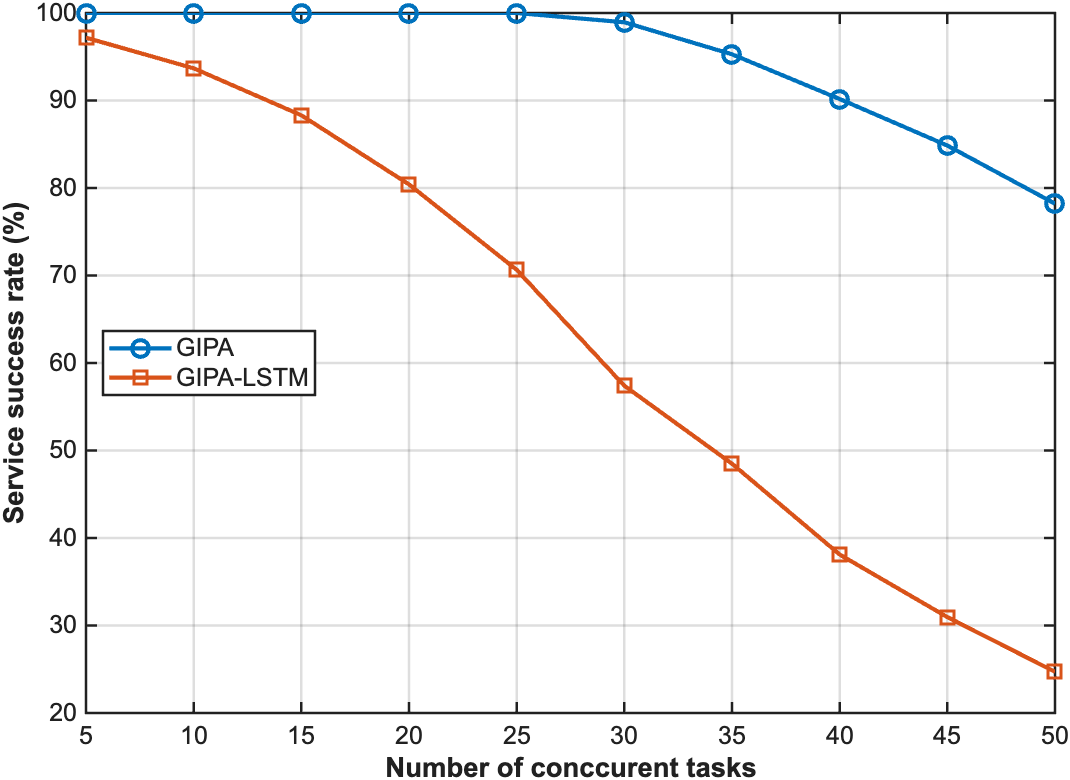}%
\caption{Performance of several methods when orchestrating navigation services.}
\end{figure}
To verify the superior advantage of the GDM in intent prediction compared to traditional models, we trained an intent prediction model based on the LSTM architecture for use in GIPA's intent prediction and observed its service execution performance compared to the standard GIPA. As shown in Fig. 6, the performance curve of the LSTM-based GIPA is significantly lower than that of the GDM-based GIPA, and the performance decays more rapidly as the number of tasks increases. This is because user implicit intent exhibits high randomness and complex distribution. As a regression model, LSTM tends to predict the mean of new data based on historical data. It is often unsuitable for the highly complex intents of edge network users, where, for instance, the same explicit intent corresponds to different resource requirements in various scenarios. Through the reverse denoising process and random noise, GDM possesses a strong exploration capability, allowing it to capture the diversity of the intent distribution. This enables GIPA to generate more precise and diverse implicit intent vectors, thereby guiding the downstream PSOM to make superior orchestration decisions.

\section{Conclusion}
In this paper, to address the challenges posed by high user mobility and the implicit nature of service intent in edge networks, we have proposed a SFC orchestration framework empowered by a Generative Intent Prediction Agent. First, we have constructed a multi-dimensional intent space model that quantifies vague natural language instructions into concrete constraints across three dimensions: functionality, Quality of Service (QoS), and resource requirements. To address the high randomness and non-linear characteristics of user intent evolution, we have designed an implicit intent prediction method based on GDM to accurately reconstruct users' latent demands from complex spatio-temporal contexts. Furthermore, we have embeded the predicted implicit intent vector into the SFC orchestration model, guiding the network to proactively optimize the SFC deployment strategy based on future potential demands. Experimental results have demonstrated that GIPA significantly outperforms baseline methods in high-concurrency and high-dynamic scenarios. Particularly in service scenarios involving user mobility, GIPA has significantly reduced end-to-end delay and maximized the service success rate by proactively sensing user intent and actively migrating services.

\bibliography{reference}
\bibliographystyle{IEEEtran}

\begin{IEEEbiography}[{\includegraphics[width=1in,height=1.25in,clip,keepaspectratio]{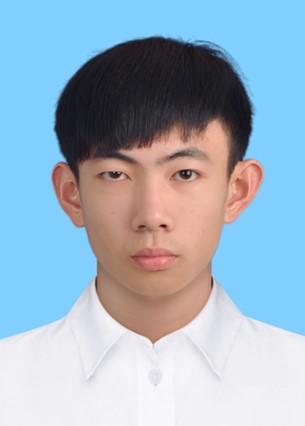}}]{Yan Sun}
is currently pursuing his Ph.D. degree in the State Key Laboratory of Networking and Switching Technology, Beijing University of Posts and Telecommunications, Beijing, China. His main research interests are intent-driven network, large AI model and edge computing. E-mail: sunyan79@bupt.edu.cn.
\end{IEEEbiography}
\begin{IEEEbiography}[{\includegraphics[width=1in,height=1.25in,clip]{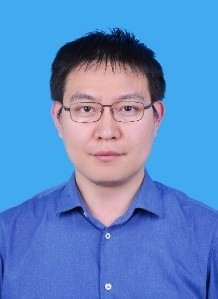}}]{Shaoyong Guo}
is currently a professor with the State Key Laboratory of Networking and Switching Technology, Beijing University of Posts and Telecommunications. He received the National Science Fund for Excellent Young Scholars in 2023. His research interests include DPU, Blockchain Application technology, Edge Intelligence, and so on. He has achieved innovative results such as power communication network convergence control model and method, network data trusted sandbox privacy sharing service mechanism and technology, and edge security protection technology and mechanism in an open network environment. He is undertaking many key research and development projects and fund projects, and contributed to a number of pioneering standards proposals in ITU-T. The systems and devices developed by him have large-scale application. He was awarded the second prize of science and technology progress in Beijing and Henan province respectively, the second prize of Science and Technology Progress Award of Chinese Institute of Electronics, and so on. E-mail: syguo@bupt.edu.cn.
\end{IEEEbiography}

\begin{IEEEbiography}[{\includegraphics[width=1in,height=1.25in,clip,keepaspectratio]{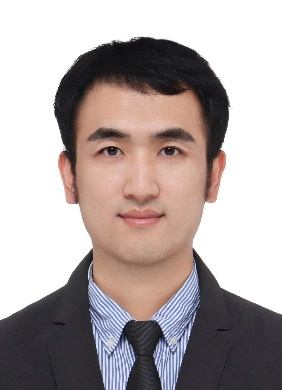}}]{Sai Huang}
, is currently working at the Department of Information and Communication Engineering as a professor, Beijing University of Posts and Telecommunications, and serves as the academic secretary of the Key Laboratory of Universal Wireless Communications, Ministry of Education, P.R. China. He is the IEEE senior member and the reviewer of international journals such as IEEE Transactions on Wireless Communications, IEEE Transactions on Vehicular Technology, IEEE Wireless Communications Letters, IEEE Transactions on Cognitive Communications and Networking and International Conferences such as IEEE ICC and IEEE GLOBECOM. His research directions are machine learning assisted intelligent signal processing, statistical spectrum sensing and analysis, fast detection and depth recognition of universal wireless signals, millimeter wave signal processing and cognitive radio network.\end{IEEEbiography}
\begin{IEEEbiography}[{\includegraphics[width=1in,height=1.25in,clip,keepaspectratio]{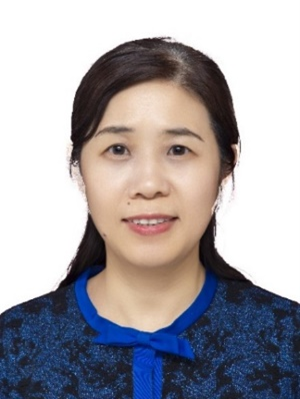}}]{Zhiyong Feng}
, is a senior member of IEEE and a full professor. She is the director of the Key Laboratory of Universal Wireless Communications, Ministry of Education. She holds B.S., M.S., and Ph.D. degrees in Information and Communication Engineering from Beijing University of Posts and Telecommunications (BUPT), Beijing, China. She is a technical advisor of NGMN, the editor of IET Communications, and KSII Transactions on Internet and Information Systems, the reviewer of IEEE TWC, IEEE TVT, and IEEE JSAC. She is active in ITU-R, IEEE, ETSI and CCSA standards. Her main research interests include wireless network architecture design and radio resource management in 5th generation mobile networks (5G), spectrum sensing and dynamic spectrum management in cognitive wireless networks, universal signal detection and identification, and network information theory.\end{IEEEbiography}
\begin{IEEEbiography}[{\includegraphics[width=1in,height=1.25in,clip]{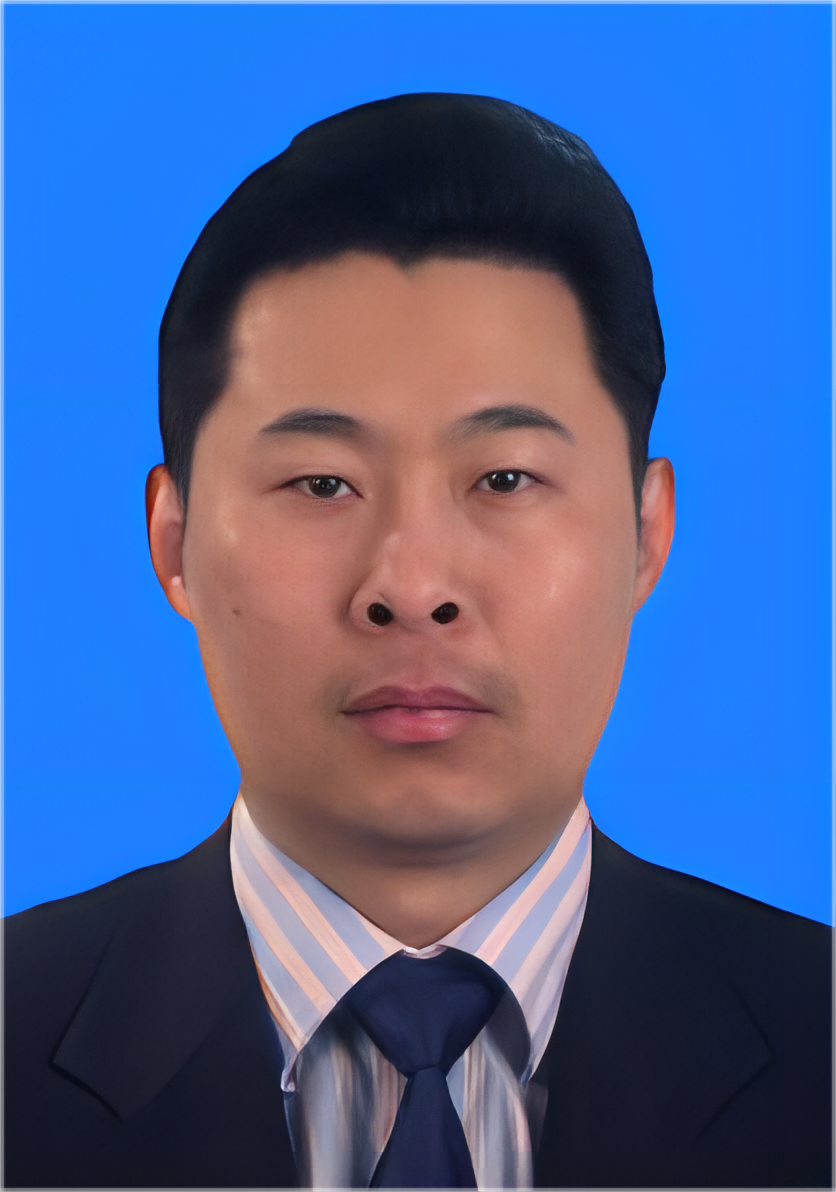}}]{Feng Qi}
is currently a Professor with the State Key Laboratory of Networking and Switching Technology, Beijing University of Posts and Telecommunications, Beijing, China, and a Researcher with Peng Cheng Laboratory, Shenzhen, China. His research interests include communications software, network management, and business intelligence. He has won two National Science and Technology Progress Awards. He has also written more than 10 ITU-T international standards and industry standards. He was the Vice Chairman of ITU-T Study Group 4 and Study Group 12. E-mail: qifeng@bupt.edu.cn.
\end{IEEEbiography}
\begin{IEEEbiography}
[{\includegraphics[width=1in,height=1.25in,clip,keepaspectratio]{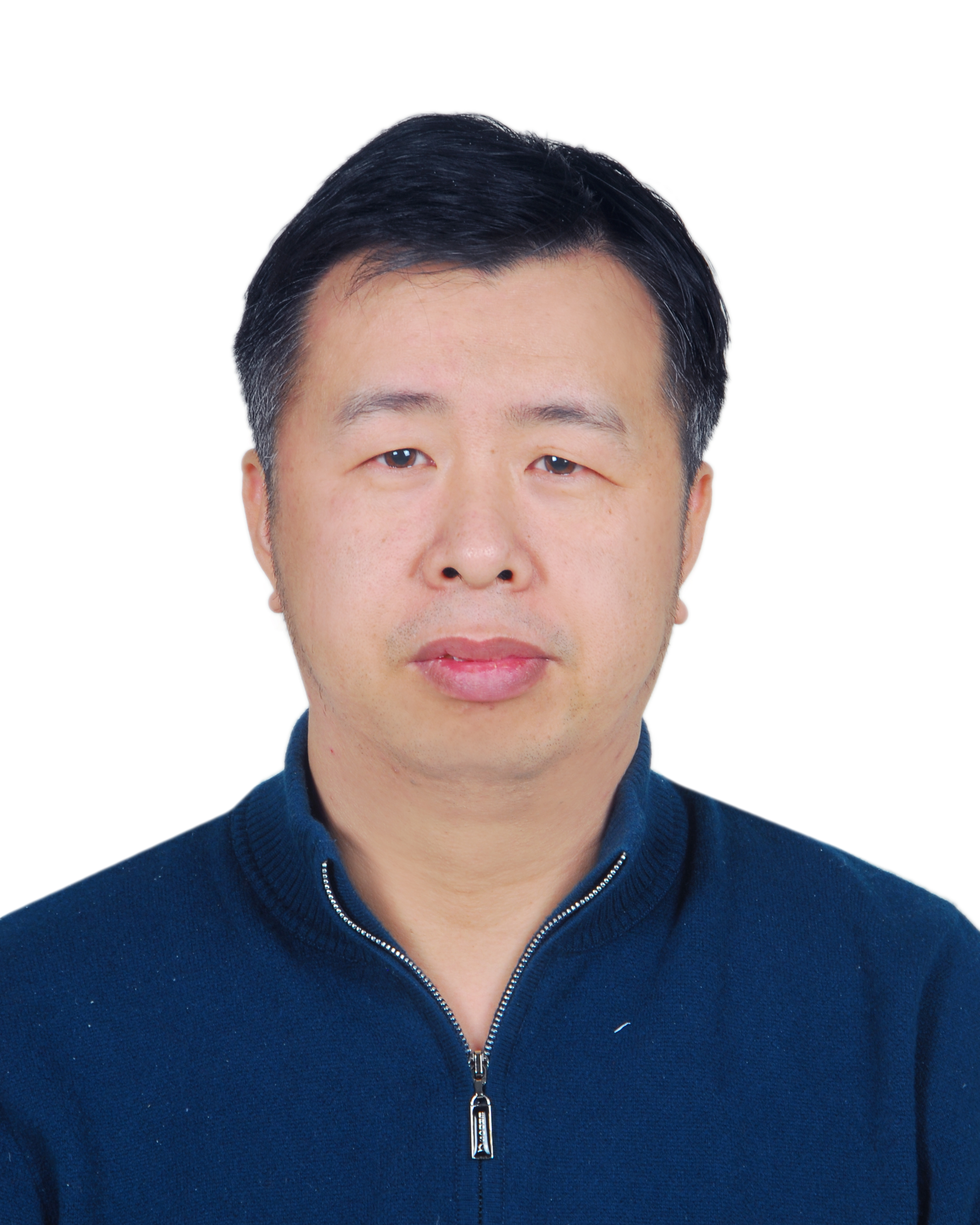}}]{Xuesong Qiu}
(Senior Member, IEEE) received the Ph.D. degree from the Beijing University of Posts and Telecommunications, Beijing, China, in 2000. He is currently a Professor and the Ph.D. Supervisor with the State Key Laboratory of Networking and Switching Technology, Beijing University of Posts and Telecommunications. He has authored about 100 SCI/EI index papers. He presides over a series of key research projects on network and service management, including the projects supported by the National Natural Science Foundation and the National HighTech Research and Development Program of China. E-mail: xsqiu@bupt.edu.cn.
\end{IEEEbiography}

\end{document}